\documentclass[twocolumn,prl,aps]{revtex4}

\begin{document}
\input psfig
\pssilent

\title{Canonical quantization of general relativity in discrete space-times}

\author{Rodolfo Gambini$^1$, and Jorge Pullin$^2$}

\affiliation{1. Instituto de F\'{\i}sica, Facultad de Ciencias,
Universidad
de la Rep\'ublica, Igu\'a 4225, CP 11400 Montevideo, Uruguay\\
2. Department of Physics and Astronomy, Louisiana State
University, 202 Nicholson Hall, Baton Rouge, LA 70803}

\date{October 27th 2002}

\begin{abstract}
It has long been recognized that lattice gauge theory formulations,
when applied to general relativity, conflict with the invariance of
the theory under diffeomorphisms. Additionally, the traditional
lattice field theory approach consists in fixing the gauge in a
Euclidean action, which does not appear appropriate for general
relativity.  We analyze discrete lattice general relativity and
develop a canonical formalism that allows to treat constrained
theories in Lorentzian signature space-times.  The presence of the
lattice introduces a ``dynamical gauge'' fixing that makes the
quantization of the theories conceptually clear, albeit
computationally involved. Among other issues the problem of a
consistent algebra of constraints is automatically solved in our
approach. The approach works successfully in other field theories as
well, including topological theories like BF theory.  We discuss a
simple cosmological application that exhibits the quantum elimination
of the singularity at the big bang.
\end{abstract}

\maketitle

Lattice approaches have been successful in Yang--Mills theory not only
in making the theory finite, but also in making it practically
computable. There has been for a long time a hope that similar
techniques could help in defining a quantization for general
relativity (see \cite{loll} for a review).  More recently,
discretizations on lattices have been used to regularize the
Hamiltonian constraint of general relativity by Thiemann \cite{Qsd}
and others and also to define the path integral in the ``spin foam''
approaches (see \cite{spinfoam} for a recent review).  A problem that
has been present since the outset is that the introduction of a
lattice breaks the symmetry of general relativity under space-time
diffeomorphisms. This has had several practical implications. For
instance, discretizations of the constraints of general relativity
fail to close an algebra like the one present in the continuum
\cite{bander}.  Moreover, it is well known that if one discretizes the
Einstein equations, the evolution equations fail to preserve the
constraints. Therefore, strictly speaking, the discrete theory is
inconsistent. This has been noted and discussed in the numerical
relativity literature \cite{choptuik}.

In this letter we would like to point out that what has been lacking
is a systematic canonical treatment of the discrete theory, both at a
classical and quantum mechanical level. When one treats the discrete
theory carefully, one notices that it has an involved canonical
structure. Dealing with it is non-trivial, since one needs a
generalization of the Dirac procedure to systems in which time is
discrete (attempting a purely spatial discretization appears highly
unnatural in the case of general relativity, where the constraints mix
space and time symmetries). We present such a generalization of the
Dirac procedure and show that the resulting theory is consistent, in
the sense that constraints do not conflict with evolution equations
and can be quantized. In our construction the lattice provides a
natural ``gauge fixing'' for the theory, which then has a ``true
Hamiltonian'' evolution. This gauge fixing may have practical use in
classical numerical relativity as well.

Let us illustrate our procedure with a mechanical system to
begin with. We take time to be discrete. The action therefore is
written as $S= \sum_{n=1}^N L(q_n,q_{n+1})$ where we
have partitioned the time interval and discretized the derivatives of
the Lagrangian. The Lagrange equations of motion are,
${\partial S / \partial q_n} = {\partial L(q_{n-1},q_n) /
\partial
q_n}+{\partial L(q_{n},q_{n+1}) / \partial q_n}=0$. In order to
introduce canonically conjugate momenta one notices that
differentiating with respect to the time derivative is equivalent to
differentiating with respect to $q_{n+1}$, or, alternatively,
$p_n = {\partial L(q_{n-1},q_n) / \partial q_n}$. The evolution of
the canonically conjugate pair $(q_n,p_n)$ can be induced through
a type 1 canonical transformation generated by,
$ F_1(q_{n},q_{n+1})\equiv -L(q_{n},q_{n+1})$. The resulting equation
of motion is $p_{n+1}={\partial L(q_{n},q_{n+1})/ \partial q_{n+1}}$,
and together with the definition of the momentum completely determines
the evolution of the system and reproduce in the limit the equations
of motion of the system \cite{DiBaGaPu}.

To quantize the resulting system we need to find a unitary operator
that implements the finite canonical transformation. Let us
particularize to the example of a particle in a potential. In such
case the discrete Lagrangian is $L(n,n+1)=m
(q_{n+1}-q_n)^2/(2\epsilon) -V(q_n)\epsilon$, where $\epsilon$ is the
time interval. The canonical transformation
yields equations of motion that can be rewritten as determining
$q_{n+1},p_{n+1}$ in terms of $q_n,p_n$. In order to quantize we need
to find a unitary operator that implements quantum mechanically
the equations of motion, $\hat{p}_{n+1}=\hat{U}
\hat{p}_n \hat{U}^\dagger$, $\hat{q}_{n+1}=
\hat{U} \hat{q}_n \hat{U}^\dagger$.
This quantization ensures that $q_n,p_n$ are such that
$[\hat{q}_n,\hat{p}_n]=i\hbar$.
The operator can be
readily found. Starting with wave functions that are function of the
coordinates $\Psi[q]$, the operator is given by, $ \hat{U}=
\exp\left({i{V(q})\epsilon\over \hbar}\right)
\exp\left({i{p}^2\epsilon\over 2m \hbar}\right)$.

We therefore see that one can set up a canonical theory for a system
with discrete time and quantize the system. What we need to do in
order to generalize this to gravity is to consider systems with
constraints. The construction proceeds as before, but the resulting
discrete equations are radically different from the continuum ones:
they are generically inconsistent {\em unless one chooses particular
values of the Lagrange multipliers}. These particular values also
make {\em the constraints automatically satisfied}: the resulting theory
is only described through evolution equations.
As before, the quantization of the theory is achieved by finding a
unitary transformation that implements the finite canonical
transformation. This is readily done for the case of the particle, at
least for simple potentials. Notice that once we solved the
constraints for the multipliers, they do not play any role in the
quantization procedure. T. D. Lee \cite{TDLee} was the first to
recognize that the discretized mechanics of a particle in a potential
could be made consistent through fixing the Lagrange multipliers (in
his language, fixing the time interval). Also Friedmann and Jack
noticed similar things in the context of Hamiltonian Regge
calculus \cite{FrJa}.

Solving the discrete constraints by choosing the Lagrange
multipliers is a procedure that is only available in the discrete
theory. This procedure might be useful in numerical relativity.
There it is well known that the discrete evolution equations (for
instance, in the ADM formulation) fail to preserve the discrete
constraints. Our proposal here consists in the following: take an
evolution step with arbitrary, unspecified, lapse and shifts. Then
impose the discrete constraints by solving the for the lapse and
shift.  There are four quantities to be determined and four
equations. The construction obviously is only guaranteed to work
locally both spatially and in time. Numerical experimentation will
be needed to decide if this is really useful in practice.

We have tested the canonical procedure with Yang--Mills theories and
BF theories in the lattice and one recovers in the former case the
traditional lattice theory. To our knowledge, our approach provides
the first lattice treatment for BF theory ever presented. Details can
be seen in \cite{DiBaGaPu}.

Let us now sketch the case of general relativity. We start by
considering the Lagrangian written in terms of Ashtekar variables,
(see for instance \cite{CDJM} and the book by Ashtekar \cite{Asbook}
page 47). To simplify treatment here we will consider the Euclidean
case, but there is no obstruction to treating the Lorentzian case, one
simply has to replace the action we are considering here with the one
introduced by Holst \cite{Holst} which leads to the real variables
of Barbero \cite{Barbero}.
The Lagrangian is,
$$
L=\int E^{ai} F_{a0}^i + \epsilon_{abc} \left[
E^{bi} E^{cj} \epsilon^{ijk} N + N^b E^{ck}\right] F_{de}^k \epsilon^{ade},
\nonumber
$$
where $E^{ai}$ and $A_a^j$ are the usual Ashtekar variables
\cite{ashtekar} and $N$ and $N^a$ are the lapse and shift. This Lagrangian
can be discretized as follows,
\begin{eqnarray}
&&L(n,n+1)= \sum_v {\rm Tr}\left[E^a_{n,v} h^a_{n,v} V_{n,v+e_1}
(h^a_{n+1,v})^\dagger (V_{n,v})^\dagger \right. \nonumber\\
&&+K_{1,n,v} h^2_{n,v} h^3_{n,v+e_2} (h^2_{n,v+e_3})^\dagger
(h^3_{n,v})^\dagger + {\rm cyc.} \nonumber\\
&&+\left.\alpha_{a,n,v} \left(h^a_{n,v} \left(h^a_{n,v}\right)^\dagger -1\right)+
\beta_{n,v} \left(V_{n,v} V^\dagger_{n,v}-1\right)\right]
\end{eqnarray}
where $K_{a,n,v} = \epsilon_{abc} \left[ E^{b}_{n,v} E^{c}_{n,v}
N_{n,v} + N^b_{n,v} E^{c}_{n,v}\right]$ and repeated indices $a,b,c$
are summed from $1$ to $3$ and ``cyc'' means cyclic permutations of
$1,2,3$.  In this expression $h^a_{n+1}$ represents
an holonomy along the $a$ direction at instant $n+1$, $V_n$ represents
the ``vertical'' (time-like) holonomy.  We will assume that the
holonomies are matrices of the form $h=h^I T^I$, $V=V^I T^I$ where
$T^0=I/\sqrt{2}$ and $T^a=-i\sigma^a/\sqrt{2}$ where $\sigma_a$ are
the Pauli matrices.  The indices $n,v$ represent a label for ``time''
$n$ and a spatial label for the vertices of the lattice $v$. The
elementary unit vectors along the spatial directions are labeled as
$e_a$, so for instance $n+e_1$ labels the nearest neighbor to $n$
along the $e_1$ direction. The quantities $E^a_{n,v}$ are elements of
the algebra of $su(2)$ and $\alpha$ and $\beta$ are Lagrange multipliers,
the last two terms of the Lagrangian enforcing the condition that the
holonomies are elements of $SU(2)$

We need to work out the equations of motion of the action. The variables
are $E^a_{n,v},h^a_{n,v},V_{n,v},\alpha^a_{n,v},\beta^b_{n,v},N_{n,v},
N^b_{n,v}$. For each variable we compute the canonical momenta at instants
$n$ and $n+1$ by $\partial L/\partial q_n$ and $\partial L/\partial q_{n+1}$
as in the examples we discussed. We will not list them explicitly
for reasons of space but we will discuss their implications. The equations
for $\alpha$ lead to a constraint that implies
the spatial holonomy $h$ is a matrix of $SU(2)$. The equations for $\beta$
lead to a similar conclusion for the holonomy $V$. The equations for $E$
lead to an evolution equation for the holonomy. The equations for $h$
lead to a constraint relating $E$ with the canonical momentum of the holonomy
at instant $n+1$ and can be viewed as an evolution equation for $E$.
The equations for $N,N^a$ give the diffeomorphism and Hamiltonian constraints
of general relativity, which are solved to determine the multipliers $N,N^a$
as we discussed above. The equations for $V$ yield Gauss' law. To
complete the canonical theory one needs to check that all constraints are
preserved in time. This can be done and it leads to a set of second class
constraint very similar to the ones we have analyzed in detail in reference
\cite{DiBaGaPu} for BF theory. One needs to introduce the Dirac brackets, which
are identical to those of BF theory and impose the constraints strongly.
This concludes the canonical formulation.

Quantization would now proceed by finding a unitary evolution operator
that implements the previous evolution equations as operator equations
quantum mechanically. Notice that there is no issue of algebra of
constraints, since the latter have been solved for the multipliers,
the lapse and shift. The procedure is computationally intensive
in a generic situation, requiring the simultaneous solution of four
non-linear coupled algebraic equations at each lattice point.  For
particular situations, the procedure might be completed in a
straightforward fashion. The study of midi-superspace situations is
therefore the next natural step in this program of quantization of
gravity. A direct application would be to study the dispersion of
waves to attempt to make contact with gamma-ray-burst phenomenology
\cite{amelino}.

For concreteness, let us exhibit the construction in a simplified
situation, that of a cosmological model. We consider a Friedmann model
with flat spatial sections coupled to a scalar field which we take to
have a large mass so we can neglect its kinetic term in the
action. This is just done in order to yield the simplest possible
cosmological model with at least one continuum degree of
freedom. Written in terms of Ashtekar's variables, the only
non-trivial quantities are one component of the triad field $E(t)$ and
one component of the connection $A(t)$ and for the scalar field its
value $\phi(t)$ and its canonically conjugate momentum $\Pi(t)$.  The
continuum Lagrangian is given by $L=E\dot{A}+\Pi \dot{\phi}
-NE^2(-A^2+\Lambda E+ m^2 \phi^2 E)$ where $N$ is the lapse (with
density weight $-1$, as is usual in the Ashtekar formulation, in terms
of the usual lapse $\alpha$ it is given by $N=E^{-3/2}\alpha$) and
$\Lambda$ is a (positive) cosmological constant. The exact solution
for this model (in the time slicing in which $\alpha=1$) has a
constant scalar field and the geometry is given by (anti)DeSitter space,
$E=\exp(2\sqrt{\Lambda+m^2 \phi^2} t)/(\Lambda+m^2 \phi^2)$,
$A=\exp(\sqrt{\Lambda+\phi^2}t)$. If we discretize the action we get,
\begin{eqnarray}
L_n&=&E_n\left(A_{n+1}-A_n\right)+\pi_n\left(\phi_{n+1}-\phi_n\right)
\nonumber\\
&&-N_n E^2_n\left(-A^2_n+\Lambda E_n+m^2 {\phi_n}^2 E_n\right)
\end{eqnarray}
and the generating function of the canonical
transformation that implements the equations of motion is given as
usual by $F_1=-L$. Working out the equations of motion and the definition
of the canonically conjugate momenta from the generating function and
the Lagrangian, one ends up with a recurrence relation for the variable
$A_n$.
\begin{equation}
A^2_{n-1}-A^2_n+2 \left( A_{n+1}-A_n\right)A_n=0\label{recurs}
\end{equation}

The lapse is completely determined by solving the Hamiltonian
constraint. This fixes dynamically the gauge: the evolution step in
time is determined by the initial conditions. For instance if we
substitute the expression for the lapse in the equations of motion one
gets, $E_n = {P^A}_{n+1} = {A_n}^2 /(\Lambda+ m^2{\phi_n}^2)$, $
A_{n+1}= A_n + \left(A^2_n-\Lambda {P^A}_n-m^2
{\phi_n}^2{P^A}_n\right)/2{A_n}$ where $P^A$ is conjugate momentum to $A$. We
see that for instance the ``step'' between $A_n$ and $A_{n+1}$ is
given by $A^2_n-\Lambda {P^A}_n-m^2 {\phi_n}^2$. Choosing the initial
values for $A$ and $P^A$ and the scalar field determines the step.

This simple model exhibits several attractive features. For instance,
for large values of the ``time label'' $n$, $A\rightarrow n^{2/3}$,
$E\rightarrow n^{4/3}$ and the lapse $N\rightarrow n^{-3}$. If we
identify $t=\epsilon n$ where $\epsilon$ is a constant, we then get
$E\rightarrow t^{4/3}$ for large times. This approximates the
continuum solution if one chooses a lapse in the continuum $\alpha\sim
t^{-1}$. We can recover any parameterization we desire in the
continuum by redefining $t$ as a function of $\epsilon n$. We
therefore see in which sense the lattice ``fixes the gauge''. Indeed,
the lattice treatment operates just like a gauge fixed treatment, but
one still has available the full reparameterization invariance of the
theory when identifying the lattice behavior with the continuum one.
Although not a traditional gauge fixing, this procedure may share with
the latter some problems. One breaks the gauge symmetries of the
theory, but it is yet to be seen, especially in complex situations if
one does not face a Gribov problem or that singularities do not appear
that prevent from covering portions of the gauge orbits, etc.  In
particular, by choosing different parameterizations we can choose to
start the region that approximates the continuum behavior as early or
as late as we desire. One can consider the recursion relation
(\ref{recurs}) and run it backwards in time. If one does that, one
finds that the universe achieves a minimum radius and the recursion
continues with the universe re-expanding again. This is a dramatically
different behavior than the one in the continuum theory. In the
latter, if one gives any initial condition and runs back in time one
will always meet an initial singularity (in this simple model, just a
coordinate singularity, in other models it is a genuine singularity
\cite{GaPupr}).  In the discrete theory one needs to fine tune the
initial data for this to happen, otherwise the singularity is avoided:
it is a point that does not fall on the grid. Quantum mechanically,
the implications of this fact are more significant. Since the
singularity is only achieved in a set of measure zero of the classical
theory, quantum mechanically one has zero probability of hitting the
singularity. Notice that this mechanism for avoiding the singularity
quantum mechanically appears distinct, although it predicts somewhat
similar results, to the one recently presented by Bojowald
\cite{bojobojo} in the context of loop quantization. Singularity
avoidance has also been discussed in other contexts \cite{Hajicek}
with different mechanisms. We would like to stress that the particular
way the singularity is avoided is model dependent and in some models
the singularity may not be avoided \cite{GaPupr}

Quantization proceeds by considering wave functions, for instance, of
the momentum ($P=P^A$) and the scalar field,
$\hat{P}\Psi(P,\phi)=P\Psi(P,\phi)$, $\hat{A} \Psi(P,\phi)=  d
\Psi(P,\phi)/dP $. One needs to implement the reality
conditions $A^*=-A$, $P^*=P$. These can be implemented via the inner
product $<\Psi|\Phi>=\int_{-\infty}^\infty d\phi  \int_{-\infty}^{\infty}
dP \Psi^*(P,\phi)\Phi(P,\phi)$. Since under this inner product $P$ has
a continuous spectrum, the probability of having a singularity (which
corresponds to $P=0$) vanishes. This can also be seen from the fact
that the expectation value of the volume is always greater than zero.
This implements in detail the behavior we had anticipated in the
heuristic discussion above.

Going back to the generic discussion of our approach, at this point
one might wonder if by taking too seriously the discrete theory we are
not neglecting the continuum limit. Two comments can be made. First of
all, there appears to be some consensus \cite{kauffman}
among researchers in quantum
gravity that the ultimate theory may have some fundamental
discreteness at the quantum level. This is in part based on the fact
that the continuum limit of Yang--Mills theories relies heavily on the
renormalizability of the theory, which appears not to be present in
the gravitational case. This is therefore motivation to take a look at
the discrete theory seriously. Moreover, having a consistent discrete
theory, which includes among its solutions some that correspond to
solutions of the continuum theory, opens the possibility of defining
an averaging procedure that could make results
discretization-independent. If one adds up solutions for various
discretizations, the solutions that have a continuum counterpart will
appear in all terms and will add up. The solutions without a continuum
counterpart will therefore be suppressed, since they will be different
for different discretizations. Notice that this allows a method of
implementing the ``sum over all discretizations'' that is advocated in
the context of spin foam models \cite{CrPeRo}.
Following the procedure outlined
above for gravity, one can construct transition amplitudes for the
discrete theory. Then, performing the averaging we just discussed for
a finite, yet large, number of discretizations will produce a result
for the transition amplitude that is approximately
discretization-independent (this is only a heuristic argument that
could face problems, for instance, discretizations could be unstable
and modes that do not correspond to the continuum limit could end up
dominating the sum). If the construction were to work one would not need
renormalizability in order to define a continuum theory in this way.
It should be noted that the general state of the art of these ideas is,
at the moment, only speculative, and there is disagreement among
various researchers.
In fact, the more conventional viewpoint for the
continuum limit of lattice gauge theories, applied to quantum
gravity does appear to work in sucessfully in $1+1$ dimensions,
and progress is being made in higher dimensionalities \cite{amb}. Even
within this more traditional viewpoint, the ``consistent'' construction
we present in this paper could play a role at the time of
selecting the correct measure of integration, since it automatically
takes into account the constraints, which is crucial at the time of
properly defining the path integral.

Another attractive element of this treatment is that the separation of
the lattice points is dynamically determined by the initial
conditions.  This implies that quantum mechanically, where generically
one has a weighted superposition of all possible initial conditions
for the system, there is an automatic ``averaging'' of all possible
discretizations per each given quantum state. It should be noted that
this averaging feature only occurs for totally constrained systems so
it is a case that diffeomorphism invariance actually helps in
getting rid of the details of the particular discretization.

Finally, the lattice treatment introduced for gravity, since it
operates as a gauge fixing, provides a solution for the ``problem
of time''. However, as it happens with usual gauge fixings one
should expect that our construction will only work locally.

Summarizing, we have presented a formulation of general relativity on
the lattice that is consistent and well defined classically and can be
readily quantized. The formulation can operate in both the Euclidean
and Lorentzian signatures and yields naturally a proposal for the
continuum limit of the theory. We exhibited its behavior in detail in
a particular cosmological situation where one sees the singularity at the big
bang disappears due to quantum effects (although this is not a generic
feature, see \cite{GaPupr}). The immediate future course of
action is to explore the formulation in more realistic,
midi-superspace models, where detailed predictions of the field theory
aspects of quantum gravity could be worked out, some of which may have
experimental implications.

We wish to thank Abhay Ashtekar, Martin Bojowald, Doug Arnold,
Alejandro P\'erez, Richard Price and Oscar Reula for comments.  This
work was supported by grants NSF-PHY0090091, funds from the Horace
Hearne Jr. Institute for Theoretical Physics and the Fulbright
Commission in Montevideo.

\end{document}